\documentclass[twocolumn]{jpsj2}
\usepackage{graphicx}
\usepackage{amsmath,amssymb}

\def\runauthor{Youichi {\sc Yanase}}

\title{ 
Antiferromagnetic Order and $\pi$-Triplet Pairing 
in the Fulde-Ferrell-Larkin-Ovchinnikov State
} 

\author{Youichi {\sc Yanase}$^{1,2,3}$\footnote{E-mail:
yanase@phys.sc.niigata-u.ac.jp} and Manfred Sigrist$^2$}

\inst{
$^{1}$ Department of Physics, University of Tokyo, Tokyo 113-0033, Japan
\\
$^{2}$ Theoretische Physik, ETH-Honggerberg, 8093 Zurich, Switzerland
\\
$^{3}$ Department of Physics, Niigata University, Niigata 950-2181, Japan 
}

\recdate{Today 2009}

\abst
{ The antiferromagnetic Fulde-Ferrell-Larkin-Ovchinnikov (AFM-FFLO) state 
of coexisting $d$-wave FFLO superconductivity and
incommensurate AFM order is studied on the basis of 
Bogoliubov-de Gennes (BdG) equations.  
 We show that the incommensurate AFM order is stabilized in the FFLO state 
by the appearance of the Andreev bound state localized around 
the zeros of the FFLO order parameter. 
 The AFM-FFLO state is further enhanced by the induced $\pi$-triplet 
superconductivity (pair density wave). 
 The AFM order occurs in the FFLO state even when it is neither stable 
in the normal state nor in the BCS state. 
 The order parameters of the AFM order, $d$-wave superconductivity, 
and $\pi$-triplet pairing are investigated by focusing on 
their spatial structures. 
 Roles of the spin fluctuations beyond the BdG equations are discussed. 
 Their relevance to the high-field superconducting phase of CeCoIn$_5$ 
is discussed. 
}

\kword
{
FFLO state, antiferromagnetism, $\pi$-triplet pairing, CeCoIn$_5$
}

\begin{document}
\sloppy
\maketitle

\newcommand{\eli}{$\acute{{\rm E}}$liashberg }
\renewcommand{\k}{\vec{k}}
\newcommand{\kk}{\vec{k'}}
\newcommand{\q}{\hspace*{0.3mm}\vec{q}\hspace*{0.5mm}}
\newcommand{\Q}{\vec{Q}}
\renewcommand{\r}{\vec{r}}
\newcommand{\e}{\varepsilon}
\newcommand{\ee}{\varepsilon^{'}}
\newcommand{\s}{{\mit{\it \Sigma}}}
\newcommand{\Tc}{$T_{\rm c}$ }
\newcommand{\Tcf}{$T_{\rm c}$}
\newcommand{\TN}{$T_{\rm N}$ }
\newcommand{\TNf}{$T_{\rm N}$}
\newcommand{\Hc}{$H_{\rm c2}^{\rm P}$ }
\newcommand{\Hcf}{$H_{\rm c2}^{\rm P}$}
\newcommand{\etal}{{\it et al.}, }
\newcommand{\PRL}{Phys. Rev. Lett. } 
\newcommand{\PRB}{{\it Phys. Rev.} B } 
\newcommand{\JPSJ}{J. Phys. Soc. Jpn. } 
\newcommand{\Science}{{\it Science} } 
\newcommand{\Nature}{{\it Nature} } 
\newcommand{\qf}{\vec{q}_{\rm F}}
\renewcommand{\i}{\hspace*{0.3mm}\vec{i}\hspace*{0.6mm}}
\renewcommand{\j}{\hspace*{0.3mm}\vec{j}\hspace*{0.6mm}}
\newcommand{\Co}{CeCoIn$_5$ }
\newcommand{\Cof}{CeCoIn$_5$}
\newcommand{\va}{\vec{a}}
\newcommand{\vb}{\vec{b}}
\newcommand{\vdelta}{\vec{\delta}\hspace*{0.5mm}}

\section{Introduction}

 The Fulde-Ferrell-Larkin-Ovchinnikov (FFLO) state in superconductors
was predicted in the 1960s by Fulde and Ferrell~\cite{FF}, and 
Larkin and Ovchinnikov~\cite{LO}. 
 In contrast to the Bardeen-Cooper-Schrieffer (BCS) state, 
Cooper pairs have a finite total momentum in the FFLO state, 
which leads to the spontaneous breaking of the spatial symmetry.  
Although this novel superconducting state with an exotic symmetry has been 
attracting much interest, the experimental search for this state 
had been fruitless for nearly 40 years. 
Under these circumstances, the discovery of a new high-field superconducting 
(HFSC) phase in \Cof~\cite{radovan2003,PhysRevLett.91.187004}, 
which is a likely candidate for the FFLO state, 
triggered many theoretical and experimental studies.~\cite{matsuda2007} 
 This recent interest on the FFLO superconductivity/superfluidity  
extends further in various related fields, such as organic 
superconductors,~\cite{uji:157001,singleton2000,lortz:187002,
shinagawa:147002,yonezawa:117002} 
cold atom gases,~\cite{Partridge01272006,Zwierlein01272006} 
astrophysics, and nuclear physics \cite{casalbuoni2004}.

 The HFSC phase of \Co has been interpreted widely within the concept of 
the FFLO state~\cite{matsuda2007,watanabe2004,capan2004,martin2005,
mitrovic2006,kumagai2006,miclea2006,correa2007,mitrovic2008,adachi2003,
ikeda:134504,ikeda:054517}. 
 However, recent observations of the AFM order in 
the HFSC phase call for a reexamination of this conclusion~\cite{young2007,kenzelmann2008,kumagaiprivate}. 
It is expected that this AFM order will be closely related to 
the AFM quantum critical point observed 
in \Cof~\cite{bianchi2003,ronning2005}. 
Therefore, the nuclear magnetic resonance 
(NMR)~\cite{young2007,kumagaiprivate} 
and neutron scattering~\cite{kenzelmann2008} 
measurements may have uncovered a novel superconducting state in 
this strongly correlated electron system. 
In particular, the neutron scattering measurement has explored 
the properties of the AFM order and found that
the wave vector of the AFM order is incommensurate 
$\q_{\rm IC} = \Q + \vdelta_{\rm IC}$ with $\vdelta_{\rm IC}$ 
along the [1,-1,0] direction for magnetic fields in the {\it ab}-plane 
of the tetragonal lattice~\cite{kenzelmann2008}. 
 The AFM staggered moment $\vec{M}_{\rm AF}$ is directed to the {\it c}-axis.

 Some scenarios have been proposed for the HFSC phase of \Co from 
the theoretical point of view. 
 We have investigated the AFM order arising from the nodal quasiparticles 
in the inhomogeneous Larkin-Ovchinnikov state~\cite{yanaseFFLOAF}. 
 The SDW order triggered by the emergence of $\pi$-triplet 
pairing~\cite{aperis2008,aperis2009,miyake2008,agterberg2009} 
has been investigated in the BCS 
state~\cite{aperis2008,aperis2009,agterberg2009} and 
in the homogeneous Fulde-Ferrell state~\cite{miyake2008}.  
 In this study, we examine the AFM-FFLO state, in which
the AFM order appears in the inhomogeneous FFLO state, 
on the basis of BdG equations. 
 The typical phase diagram in the $H$-$T$ plane and the spatial structure of 
the AFM-FFLO state are investigated in detail.

\section{Formulation}
 
 Our theoretical analysis is based on the microscopic model,
\begin{eqnarray}
  \label{eq:model}
  && \hspace{-5mm}
  H= - t \sum_{<\i,\j>,\sigma} c_{\i,\sigma}^{\dag}c_{\j,\sigma}
  + t' \sum_{<<\i,\j>>,\sigma} c_{\i,\sigma}^{\dag}c_{\j,\sigma} 
\nonumber \\ && \hspace{-0mm} 
  + U \sum_{\i} n_{{\i}\uparrow} n_{{\i}\downarrow} 
  + V \sum_{<i,j>} n_{\i} \hspace{0.5mm} n_{\j} 
\nonumber \\ && \hspace{-0mm} 
  + J \sum_{<i,j>} \vec{S}_{\i} \hspace{0.4mm} \vec{S}_{\j} 
  - g_{\rm B} \vec{H} \sum_{\i} \vec{S}_{\i}, 
\end{eqnarray}
where $\vec{S}_{\i}$ 
is the spin operator and $n_{\i}$ is the number operator at site $i$. 
 To describe the quasi-two-dimensional electronic structure 
of CeCoIn$_5$, we assume a square lattice, in which 
the bracket $<\i,\j>$ ($<<\i,\j>>$) denotes the summation over the 
nearest-neighbor sites (next-nearest-neighbor sites). 
 The on-site repulsive interaction is given by $U$, and 
$V$ and $J$ stand for the attractive and AFM  exchange interactions, 
respectively, between nearest-neighbor sites. 
 We assume $V$ to stabilize the $d$-wave superconducting state within the 
mean field BdG equations and choose $J > 0$  
for the AFM correlation in CeCoIn$_5$. 
 We assume $U=1.15$, $V=-0.4$, and $J=0.53$ throughout this paper.
 The $d$-wave superconductivity and the significant AFM spin fluctuation  
can be self-consistently described using many-body theories such as 
the fluctuation exchange (FLEX) approximation for the Hubbard or periodical 
Anderson model~\cite{yanase2003cms,yanaseFFLOQCP}.  
 We here assume the interactions $V$ and $J$ to describe these features 
in the inhomogeneous LO phase on the basis of BdG equations. 
Although the BdG equations neglect the AFM spin fluctuation 
beyond the mean field approximation, 
they are suitable for studying the qualitative features of the 
inhomogeneous superconducting and/or magnetic state. 
 We will discuss the roles of AFM spin fluctuation at the end of this paper.

 With the last term in eq.~(1), we include the Zeeman coupling due to the 
applied magnetic field. The $g$-factor is assumed to be $g_{\rm B}=2$. 
We assume the magnetic field to lie parallel to 
the [100]-axis, which we choose to be the quantized axis of spin, 
with $\vec{H} = H \hat{z}$. 
In this paper, we identify the {\it a}-, {\it b}-, and {\it c}-axes of 
the tetragonal lattice with the 
{\it z}-, {\it x}-, and {\it y}-axes for the spin, respectively. 
 We choose the unit of energy such that $t=1$ and $t'/t=0.25$. 
 The chemical potential enters as $\mu=\mu_{0} + (\frac{1}{2} U + 4 V) n_{0}$, 
where $n_{0}$ is the number density for $U=V=J=H=0$. 
 We choose $\mu_{0} = -1.05$ so as to reproduce the incommensurate 
AFM order observed in the neutron scattering measurement 
for \Cof~\cite{kenzelmann2008}. 
 Then, we obtain the number density $n \sim 0.77$.

 The BdG equations are formulated in a standard manner. 
 We take into account the Hartree term arising from $U$, $V$, and $J$. 
 The BdG equations are self-consistently solved 
for the mean fields of the spin $<\vec{S}_{\i}>$, 
charge $<n_{\i}>$, and superconductivity $\Delta_{\i,\j}^{\sigma\sigma'} = 
<c_{{\i}\sigma}c_{{\j}\sigma'}>$. 
 Since the magnetic moment is nearly opposite between the 
nearest-neighbor sites, the order parameter of the AFM order is 
described by the staggered moment defined as 
\begin{eqnarray}
  \label{eq:AFM}
  && \hspace{-10mm}
\vec{M}_{\rm AF}(\i)  = (-1)^{m+n} <\vec{S}_{\i}>,  
\end{eqnarray}
with $\i = (m,n)$. 
 Two components of the pairing field $\Delta_{\i,\j}^{\sigma\sigma'}$ 
play dominant roles in the following results. 
 The first component is the $d$-wave spin singlet pairing whose order parameter  
is described as 
\begin{eqnarray}
  \label{eq:dSC}
  && \hspace{-10mm}
\Delta^{\rm d}(\i) = \Delta_{\i,\i+\va}^{\uparrow\downarrow} 
+ \Delta_{\i,\i-\va}^{\uparrow\downarrow} 
- \Delta_{\i,\i+\vb}^{\uparrow\downarrow} - \Delta_{\i,\i-\vb}^{\uparrow\downarrow}. 
\end{eqnarray}
with $\va$ and $\vb$ being the unit vectors along 
the {\it a}- and {\it b}-axes, respectively. 
 The second component is the equal spin $\pi$-triplet pairing 
whose order parameter 
is described by the generalized $d$-vector $\vec{d}_{\rm a,b}(\i)$ as 
\begin{eqnarray}
  \label{eq:pi-triplet}
  && \hspace{-10mm}
\Delta_{\i,\i \pm \va}^{\sigma\sigma} = \pm (-1)^{m+n} 
\frac{1}{2} (-\sigma d^{\rm x}_{\rm a}(\i) + {\rm i} d^{\rm y}_{\rm a}(\i)), 
\\ && \hspace{-10mm}
\Delta_{\i,\i \pm \vb}^{\sigma\sigma} = \pm (-1)^{m+n} 
\frac{1}{2} (-\sigma d^{\rm x}_{\rm b}(\i) + {\rm i} d^{\rm y}_{\rm b}(\i)). 
\end{eqnarray}
 The $d$-wave spin singlet pairing is finite in the entire superconducting 
region, while the $\pi$-triplet pairing appears in the state of coexisting
AFM order and superconductivity~\cite{machida1981,murakami1998,Demler1995,Demler1998,aperis2008,aperis2009,miyake2008}. 
 Because of the linear coupling between the magnetic, 
spin singlet pairing, and $\pi$-triplet pairing order parameters, 
two finite order parameters among them induce the other order parameter. 
 Our model eq.~(1) includes the interactions leading to the magnetic order 
and spin singlet $d$-wave superconductivity, while the pairing interaction 
for the $\pi$-triplet pairing is negligible. 
 Therefore, the $\pi$-triplet pairing does not belong to the dominant orders, 
but is induced as a secondary order parameter by the AFM order.

 The mean field Hamiltonian is obtained as 
\begin{eqnarray}
  \label{eq:MFmodel}
  && \hspace{-10mm}
  H = -t \sum_{<\i,\j>,\sigma} c_{\i,\sigma}^{\dag}c_{\j,\sigma}
  + t' \sum_{<<\i,\j>>,\sigma} c_{\i,\sigma}^{\dag}c_{\j,\sigma} 
\nonumber \\ && \hspace{-5mm}
+\sum_{\i} (\frac{1}{2} U <n_{\i}> + V \sum_{\vdelta} <n_{\i+\vdelta}> - \mu) 
\hspace*{0.5mm} n_{\i} 
\nonumber \\ && \hspace{-5mm}
+\sum_{\i} (-2 U <\vec{S}_{\i}> + J \sum_{\vdelta} <\vec{S}_{\i+\vdelta}> 
- g_{\rm B} \vec{H}) 
\hspace*{0.5mm} \vec{S}_{\i} 
\nonumber \\ && \hspace{-5mm}
+ \frac{1}{2}  \sum_{\i,\vdelta,\sigma,\sigma'} [\Delta_{\i,\i+\vdelta}^{\sigma\sigma'} 
  \hspace*{0.5mm} c_{\i,\sigma}^{\dag} c_{\i+\vdelta,\sigma'}^{\dag}  + c.c.], 
\end{eqnarray}
where the summation of $\vdelta $ is taken over 
$\vdelta = \pm \va,\pm \vb$. 
 The pairing field is obtained as 
$\Delta_{\i,\i+\vdelta}^{\uparrow\downarrow} = 
-(V - J/4) <c_{\i,\uparrow} c_{\i+\vdelta,\downarrow}>
- J/2 <c_{\i,\downarrow} c_{\i+\vdelta,\uparrow}>$ 
and 
$\Delta_{\i,\i+\vdelta}^{\sigma\sigma} = 
-(V + J/4) <c_{\i,\sigma} c_{\i+\vdelta,\sigma}>$. 
 The thermodynamic average $<>$ is obtained on the basis of the 
mean field Hamiltonian eq.~(\ref{eq:MFmodel}). 
 The self-consistent equations for $<\vec{S}_{\i}>$, $<n_{\i}>$, and 
$\Delta_{\i,\i+\vdelta}^{\sigma\sigma'}$ for each $\i$ yield the BdG equations.

 The free energy is obtained as  
\begin{eqnarray}
  \label{eq:Freeenergy}
  && \hspace{-10mm}
  F = - \frac{1}{2} T \sum_{\alpha} \log [1 + \exp (-E_{\alpha}/T)] 
\nonumber \\ && \hspace{-5mm}
+ \sum_{\i} (\frac{1}{4} U <n_{\i}>^{2} 
+ \frac{1}{2} V \sum_{\vdelta} <n_{\i}> <n_{\i+\vdelta}>)
\nonumber \\ && \hspace{-5mm}
+ \sum_{\i} (-U <\vec{S}_{\i}>^{2}
+ \frac{1}{2} J \sum_{\vdelta} <\vec{S}_{\i}> \cdot <\vec{S}_{\i+\vdelta}> ) 
\nonumber \\ && \hspace{-5mm}
+ \frac{1}{2} \sum_{\i,\vdelta,\sigma,\sigma'} \Delta_{\i,\i+\vdelta}^{\sigma\sigma'} 
<c_{\i+\vdelta,\sigma'}^{\dag} c_{\i,\sigma}^{\dag}>, 
\end{eqnarray}
where $E_{\alpha}$ is the energy of Bogoliubov quasiparticles obtained by 
the mean field Hamiltonian eq.~(\ref{eq:MFmodel}). 
 The stable phase is determined by minimizing the free energy for the 
self-consistent solutions of the BdG equations.

\section{Results}

\subsection{Phase diagram}

 We first show the phase diagram against the temperature and magnetic field. 
 The normal, uniform BCS, purely FFLO, and AFM-FFLO states are shown in Fig.~1. 
At high magnetic fields, the phase transition from the normal state to 
a superconducting state is of the first order, 
consistent with the experimental results for 
\Cof~\cite{matsuda2007,bianchi2002,tayama2002}. 
 We have shown that both the on-site repulsion $U$ and AFM 
interaction $J$ are required to reproduce the first order phase transition 
to the FFLO state~\cite{yanaseFFLOdisorder}. 
 This indicates that both the AFM spin fluctuation and the local electron 
correlation play an essential role in the phase diagram of \Cof. 
These features are qualitatively understood on the basis of  
the Fermi liquid theory~\cite{vorontsov2006}.  
 Although the size of the numerical calculation ($40 \times 40$ lattices)
is not sufficiently large for distinguishing the first and second order 
phase transitions from the BCS state to the FFLO state (BCS-FFLO transition), 
the BCS-FFLO transition is expected to be of the second order, 
as shown for a similar model~\cite{yanaseFFLOAF}. 
 The second order BCS-FFLO transition is described by the 
nucleation of the FFLO nodal 
plane~\cite{machida1984,yanaseFFLOAF,vorontsov2006}. 
Because of our tractable system size limitation, maximally 
two nodal planes fit into our calculation, as shown in Fig.~2.

 An important finding obtained from Fig.~1 is the appearance of 
the AFM-FFLO state.  
 The Ne\'el temperature $T_{\rm N}$ in the BCS and normal states is 
less than $0.01$, which is the lower limit of our calculation, 
while we obtain a much higher Ne\'el temperature $T_{\rm N}$ of $ \sim 0.02$ 
in the FFLO state. 
 Thus, the AFM order is favored in the FFLO state rather than 
in the BCS and normal states.

\begin{figure}[h]
\begin{center}
\includegraphics[width=7cm]{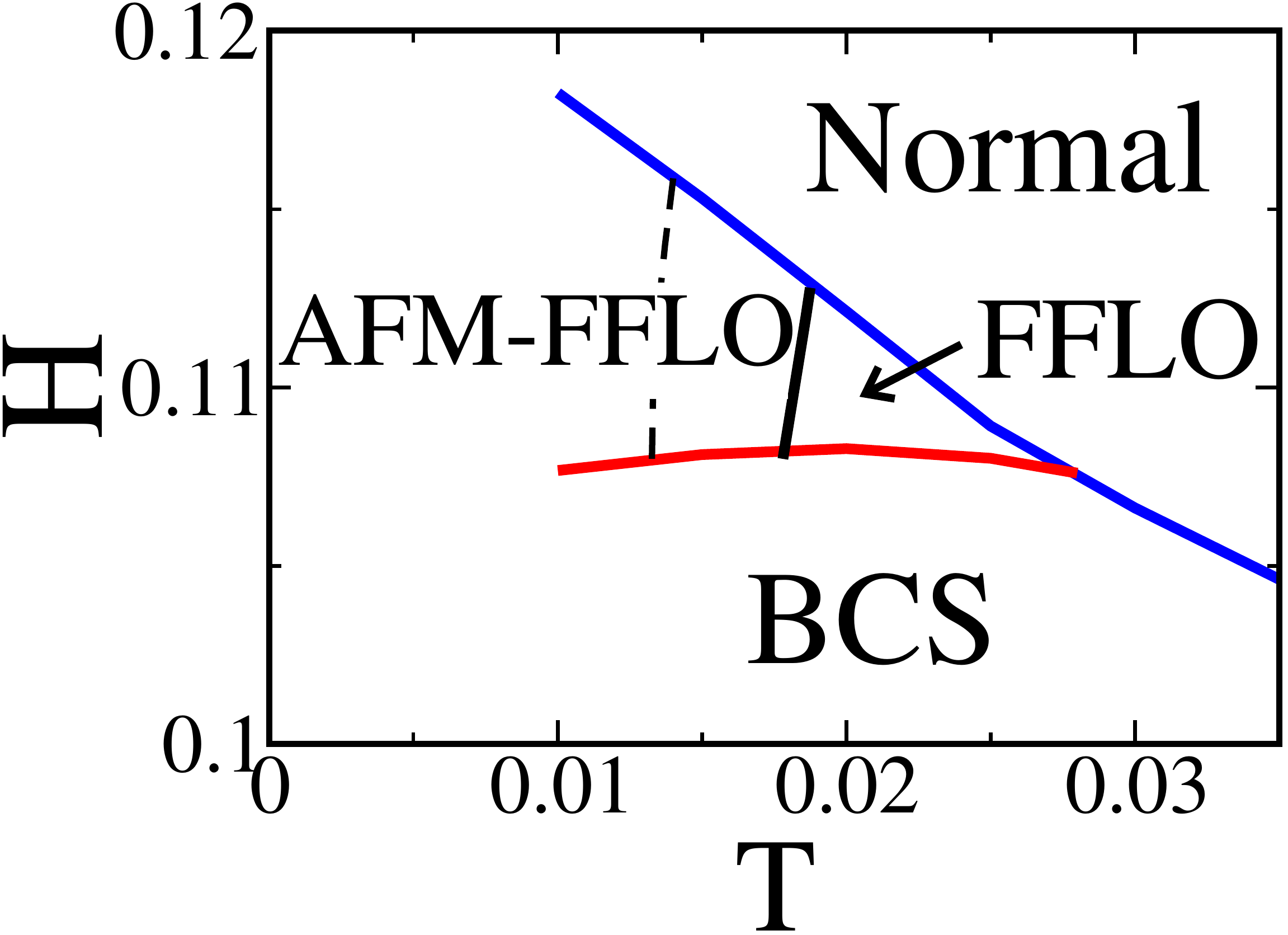}\hspace{1pc}%
\caption{(Color online)
Phase diagram for the magnetic field and temperature. 
The normal, uniform BCS, FFLO, and AFM-FFLO states are 
shown in the figure. 
Solid lines show the phase boundary between these states. 
The thin dashed line shows the fictitious phase transition line 
from the FFLO state to the AFM-FFLO state 
where the order parameter of $\pi$-triplet pairing 
is neglected. 
Note that the \Tc of superconductivity at $H=0$ is $T_{\rm c} = 0.096$.
}
\end{center} 
\end{figure}

 Two mechanisms stabilize the AFM order in the FFLO state. 
 The first mechanism is the appearance of Andreev bound states around 
the spatial nodes of the modulated superconducting order parameter 
in the FFLO state. 
 The $\pi$-phase shift of the order parameter introduced in the FFLO state 
produces a large local density of states (DOS) and 
triggers the AFM order~\cite{yanaseFFLOAF}. 
 This is the reason why the AFM order is favored in the FFLO state 
rather than in the normal and BCS states. 
 The appearance of the Andreev bound states is a characteristic feature 
of the inhomogeneous Larkin-Ovchinnikov state and is absent 
in the homogeneous Fulde-Ferrell state. 
 The role of the Andreev bound states in the FFLO nodal structure 
is much more pronounced than that of the single vortex that has been 
investigated by Ogata~\cite{ogata1999}. 
 This is because the FFLO nodal plane is a two-dimensional object, 
while the vortex is an one-dimensional object.

 The second mechanism is the linear coupling between the AFM order and 
the $\pi$-triplet pairing that favors the AFM order in the spin singlet 
superconducting state. 
 It has been shown that this coupling stabilizes the AFM state 
in the BCS state~\cite{aperis2008,aperis2009} and that in the 
Fulde-Ferrell state\cite{miyake2008}. 
 In contrast to the previous studies~\cite{aperis2008,aperis2009}, 
the $\pi$-triplet pairing state without the AFM order 
is hardly stabilized in our model. 
 However, the coupling to the $\pi$-triplet pairing significantly stabilizes 
the AFM order in the Larkin-Ovchinnikov state. 
 The thin dashed line in Fig.~1 shows the fictitious Ne\'el temperature 
in the FFLO state, if the $\pi$-triplet pairing is neglected. 
 We observe the substantial increase in $T_{\rm N}$ due to the admixed 
$\pi$-triplet pairing. 
 Thus, the $\pi$-triplet pairing plays a quantitatively important role 
in the stability of the AFM-FFLO state even when the amplitude of 
$\pi$-triplet pairing is small.

 We now turn to the question why the AFM order appears discontinuously 
at the normal-FFLO transition as well as at the BCS-FFLO transition. 
 The former discontinuity is simply caused by the first order 
phase transition to the FFLO state. 
The order parameter of $d$-wave superconductivity appears discontinuous 
at the normal-FFLO transition and the AFM state is directly coupled 
to this order. 
 The discontinuity at the BCS-FFLO transition looks more surprising 
because the phase transition is of the second order. 
 The key lies in the appearance of the Andreev bound states 
that we have mentioned above. 
 The second order phase transition from the BCS state to the FFLO state 
is associated with the nucleation of domain walls in the superconducting order 
parameter~\cite{machida1984,yanaseFFLOAF,vorontsov2006}. 
 The Andreev bound states localized around such domain walls are the 
source of the AFM instability. 
 Since the Andreev bound states in the isolated domain wall 
are nearly independent of the adjacent domain walls, within 
the mean field approach, the Ne\'el temperature is nearly independent 
of the density of domain walls. 
 In other words, the AFM order occurs immediately when domain walls 
are nucleated at the BCS-FFLO transition~\cite{yanaseFFLOAF}. 
 Note that the spatially averaged magnetic moment is continuous 
at the second order BCS-FFLO transition, although the Ne\'el temperature 
is discontinuous there. 
 The phase boundary near the BCS-FFLO transition obtained here 
by the BdG equations would be altered by the spin fluctuation, 
as discussed in \S4. 
 However, the spin fluctuations do not alter the result that 
the AFM order is confined in the FFLO state. 
 This feature of the phase diagram is consistent with 
the  experiment of \Co that revealed the magnetic order in the 
HFSC phase, but neither in the normal state nor in the low field 
superconducting phase~\cite{kenzelmann2008}.

\subsection{Spatial structure of AFM-FFLO state}

 We here investigate the spatial structures of the AFM-FFLO state 
shown in Fig.~1. 
 Figure~2(a) depicts the order parameter of the $d$-wave spin singlet 
superconductivity $\Delta^{\rm d}(\i)$, while Fig.~2(b) shows 
the AFM staggered moment $M^{\rm x}_{\rm AF}(\i)$ for $\i = (m,n)$. 
 In Fig.~2(a), we assume that the modulation vector of FFLO superconductivity 
is parallel to the magnetic field $\q_{\rm FFLO} \parallel \vec{H}$,  
in accordance with ref.~22. 
 We find that the AFM staggered moment is perpendicular to the applied 
magnetic field, consistent with the neutron scattering 
measurement~\cite{kenzelmann2008}. 
 Since we neglect the spin-orbit coupling, the AFM state with 
$\vec{M}_{\rm AF}(\i) \parallel \hat{x}$ is degenerate with that 
with $\vec{M}_{\rm AF}(\i) \parallel \hat{y}$. We choose the former 
in the following results.

\begin{figure}[h]
\begin{center}
\vspace*{-0mm}
\hspace*{-20mm}
\includegraphics[width=10cm]{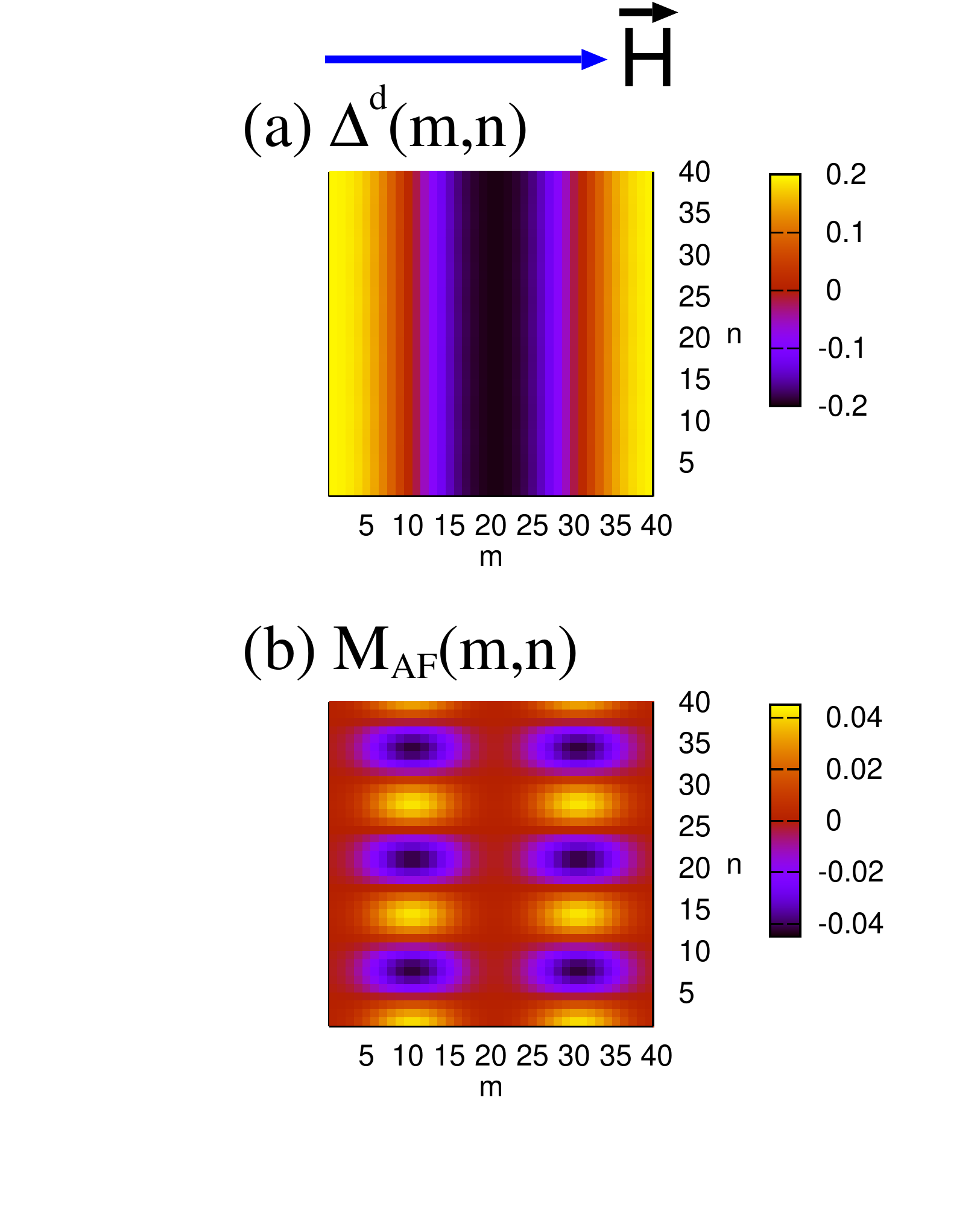}\hspace{2pc}%
\vspace*{-15mm}
\caption{(Color online)
Spatial dependence of (a) the spin singlet pairing field with 
$d$-wave symmetry $\Delta^{\rm d}(\i)$ and 
(b) the AFM staggered moment $M^{\rm x}_{\rm AF}(\i)$ for $\i = (m,n)$. 
We assume $T=0.01$ and $H=0.112$. 
The direction of the magnetic field is shown by an arrow. 
}
\end{center}
\end{figure}

 We observe that the AFM staggered moment is localized around 
the spatial nodes of the modulated order parameter 
of $d$-wave superconductivity. 
 This is because the Andreev bound states mainly induce the AFM order, 
as previously mentioned. 
 Another intriguing finding is the direction of the 
incommensurate wave vector, which is perpendicular to 
the FFLO modulation vector as 
$\vdelta_{\rm IC} \perp \q_{\rm FFLO} \parallel \vec{H}$. This structure is 
consistent with the experimental result of \Cof~\cite{kenzelmann2008}. 
 The amplitude of the incommensurate wave vector $|\vdelta_{\rm IC}|$ is 
independent of the density of FFLO nodal planes, 
as we have shown in ref.~30. 
 Therefore, our results are compatible with the experimental observation 
in which $\q_{\rm IC} = \Q + \vdelta_{\rm IC}$ is independent of  
magnetic field~\cite{kenzelmann2008}.   
 Thus, our results on the direction, spatial structure, and 
magnetic field dependences of the AFM staggered moment 
are consistent with the neutron scattering measurement~\cite{kenzelmann2008}.

\begin{figure}[h]
\begin{center}
\vspace*{-7mm}
\hspace*{-20mm}
\includegraphics[width=10cm]{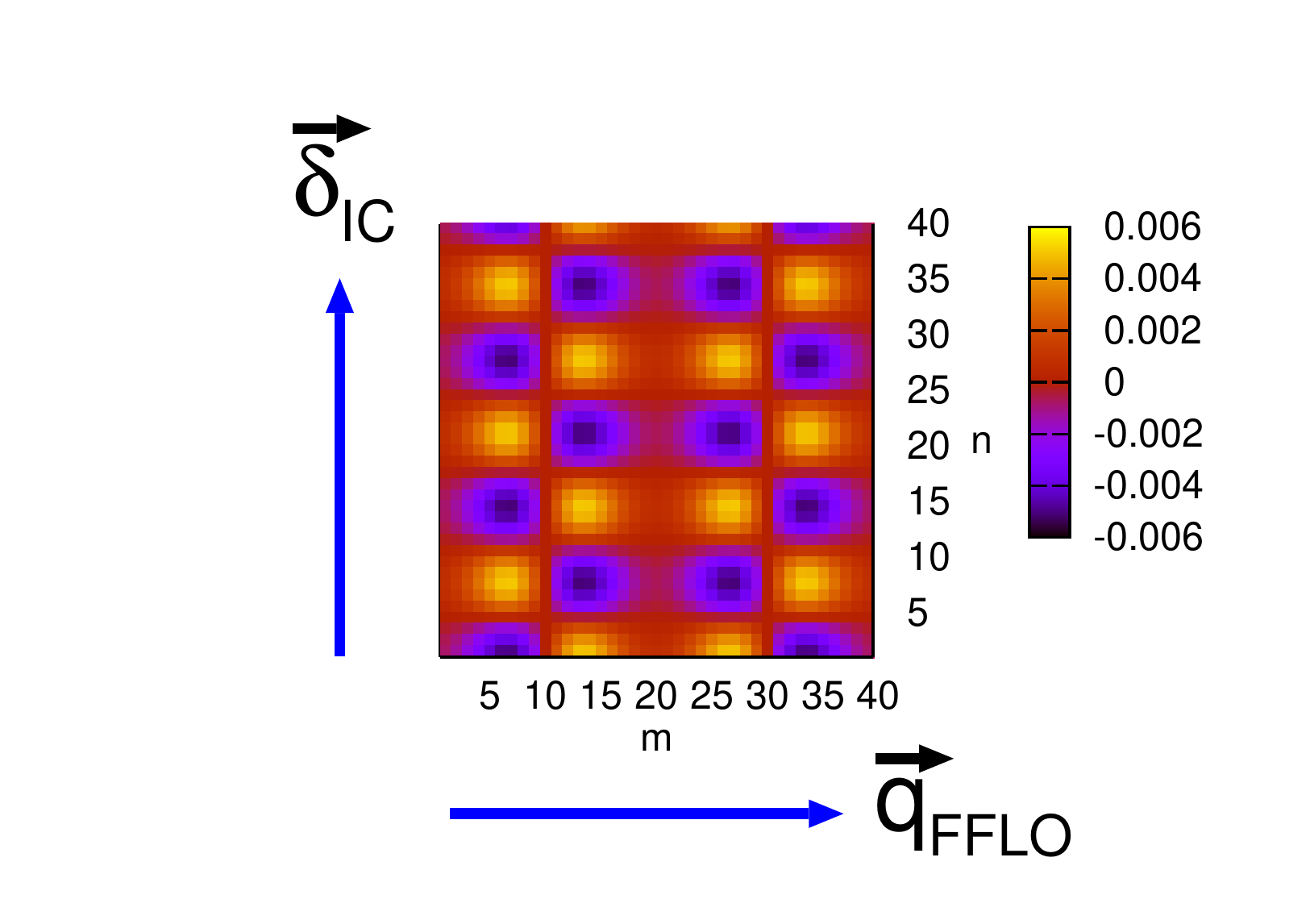}\hspace{2pc}%
\vspace*{-10mm}
\caption{(Color online)
Spatial dependence of the order parameter 
for the $\pi$-triplet pairing.  
We plot 
$-d^{\rm x}_{\rm a}(\i) + {\rm i} d^{\rm y}_{\rm a}(\i) = (-1)^{m+n}
(\Delta_{\i,\i + \va}^{\uparrow\uparrow} - \Delta_{\i,\i - \va}^{\uparrow\uparrow})$. 
The directions of the FFLO modulation $\q_{\rm FFLO}$ and 
incommensurability $\vdelta_{\rm IC}$ are shown by arrows. 
The parameters are the same as those in Fig.~2. 
}
\end{center}
\end{figure}

 Figure~3 shows the order parameter of the $\pi$-triplet pairing. 
 We obtain the generalized $d$-vector as $\vec{d}_{\rm a}(\i) \simeq 
- \vec{d}_{\rm b}(\i) \propto \hat{x} + {\rm i} \alpha \hat{y}$ with 
$0 < \alpha < 1 $. 
 This structure is the same as the $d$-vector proposed for 
the high-field superconducting phase in Sr$_2$RuO$_4$~\cite{udagawa2005}. 
 Since $d^{\rm y}_{\rm a,b}(\i)$ is pure imaginary for any $\i$, 
we plot the real quantity 
$-d^{\rm x}_{\rm a}(\i) + {\rm i} d^{\rm y}_{\rm a}(\i)$ in Fig.~3. 
 The complex spatial structure of the $\pi$-triplet pairing arises 
from the spatial modulation in the AFM staggered moment 
$M^{\rm x}_{\rm AF}(\i)$ along the {\it b}-direction and 
that in the spin singlet pairing field $\Delta^{\rm d}(\i)$ along the 
{\it a}-direction. 
 Since the $\pi$-triplet pairing is induced by the combination of 
the AFM order and spin singlet pairing, the generalized $d$-vector for the 
$\pi$-triplet pairing $\vec{d}_{\rm a,b}(\i)$ changes the sign 
at the zeros of $M^{\rm x}_{\rm AF}(\i)$ and those of $\Delta^{\rm d}(\i)$. 
 The modulation wave vectors of the $\pi$-triplet pairing 
are mainly $\q_{\rm FFLO}$ along the {\it a}-axis and $\vdelta_{\rm IC}$ 
along the {\it b}-axis.

\begin{figure}[h]
\begin{center}
\vspace*{-16mm}
\hspace*{-20mm}
\includegraphics[width=10cm]{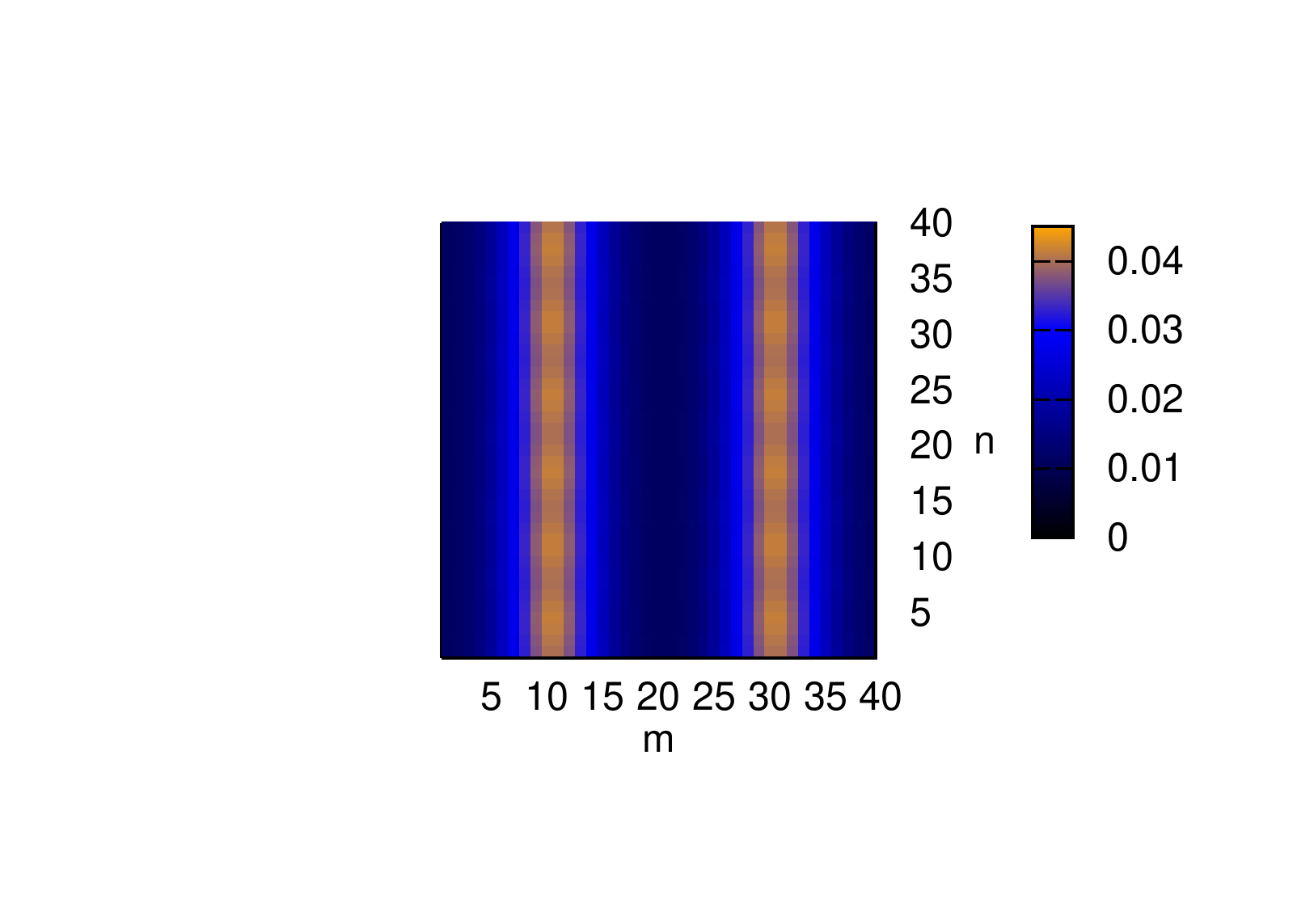}\hspace{2pc}%
\vspace*{-15mm}
\caption{(Color online)
Spatial dependence of the magnetization 
$M^{\rm z}(\i) = <S^{\rm z}_{\i}>$ parallel to the magnetic field. 
The parameters are the same as those in Fig.~2. 
}
\end{center}
\end{figure}

Finally, we show the spatially inhomogeneous
magnetization along the magnetic field $M^{\rm z}(\i)$ (Fig.~4). 
 The magnetization mainly arises from the polarization of 
the Andreev bound states~\cite{ichioka2007}. 
 Therefore, we clearly observe the magnetization localized around the 
spatial nodes of the spin singlet pairing field. 
 In addition, a weak modulation of magnetization appears 
along the nodal lines owing to the incommensurate AFM order. 
 It is shown that the magnetization is enhanced at the intersection points 
between the nodal planes of the incommensurate AFM order and the 
spin singlet pairing.

\subsection{Neutron scattering}

 We here propose an experiment that can unambiguously identify the 
AFM-FFLO state. 
 Figure~5 shows the Fourier transformation of the magnetic moment 
perpendicular to the magnetic field, 
\begin{eqnarray}
  \label{eq:ssfactor}
  && \hspace{-10mm}
S^{\rm x}(\q) = \frac{1}{N} \sum_{\i} <S^{\rm x}_{\i}> \exp({\rm i} \hspace*{0.3mm}
\q \cdot \i). 
\end{eqnarray}
\begin{figure}[h]
\begin{center}
\vspace*{-15mm}
\hspace*{-20mm}
\includegraphics[width=10cm]{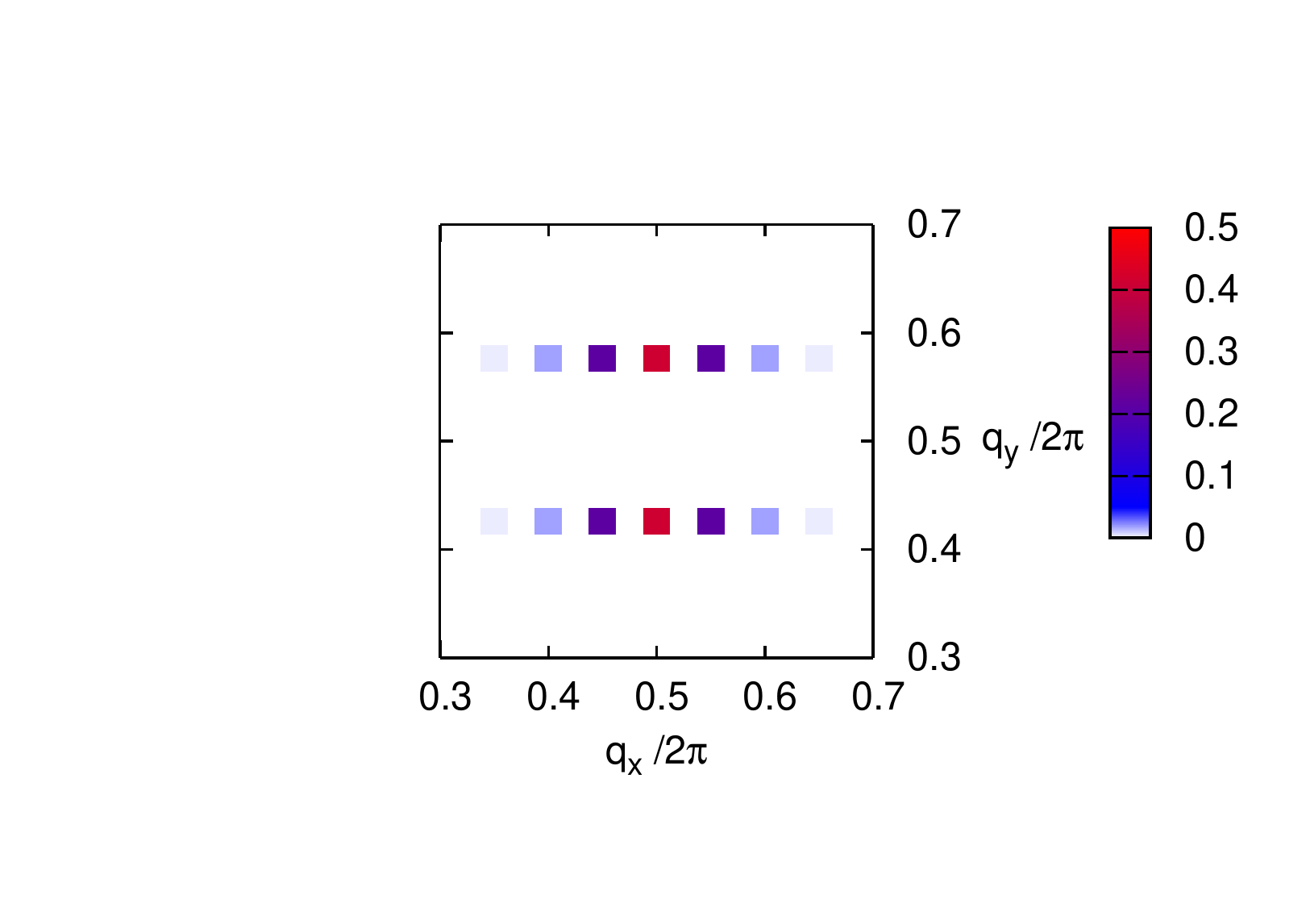}\hspace{2pc}%
\vspace*{-15mm}
\caption{(Color online)
$S^{\rm x}(\q)$ in the AFM-FFLO state for $\q = (q_{\rm x},q_{\rm y})$. 
The parameters are the same as in Fig.~2. 
}
\end{center}
\end{figure}
 We observe two sharp peaks in $S^{\rm x}(\q)$ at the incommensurate 
wave vectors $\q_{\rm IC} = \Q \pm \vdelta_{\rm IC}$, 
where $\vdelta_{\rm IC}/2\pi = (0, 0.075)$ in our calculation. 
 In addition to these main peaks, the satellite peaks appear 
at $\q = \q_{\rm IC} + \q_{\rm st}$ with 
$\q_{\rm st} \parallel \vec{H} \perp \vdelta_{\rm IC}$ in the AFM-FFLO state. 
 These satellite peaks are the characteristic feature of the AFM-FFLO state 
and unambiguously show the spontaneous translation symmetry breaking 
along the magnetic field. 
 Therefore, if the spin structure in Fig.~5 is observed by a neutron 
scattering measurement, that would give unambiguous evidence for 
the AFM order in the FFLO state. 
 The  neutron scattering measurements have revealed a sharp peak 
at $\q = \q_{\rm IC}$~\cite{kenzelmann2008}, 
however, the presence of satellite peaks shown in Fig.~5 has not yet 
been explored experimentally.

\section{Roles of Spin Fluctuation}

\begin{figure}[h]
\begin{center}
\hspace*{6mm}
\includegraphics[width=7.5cm]{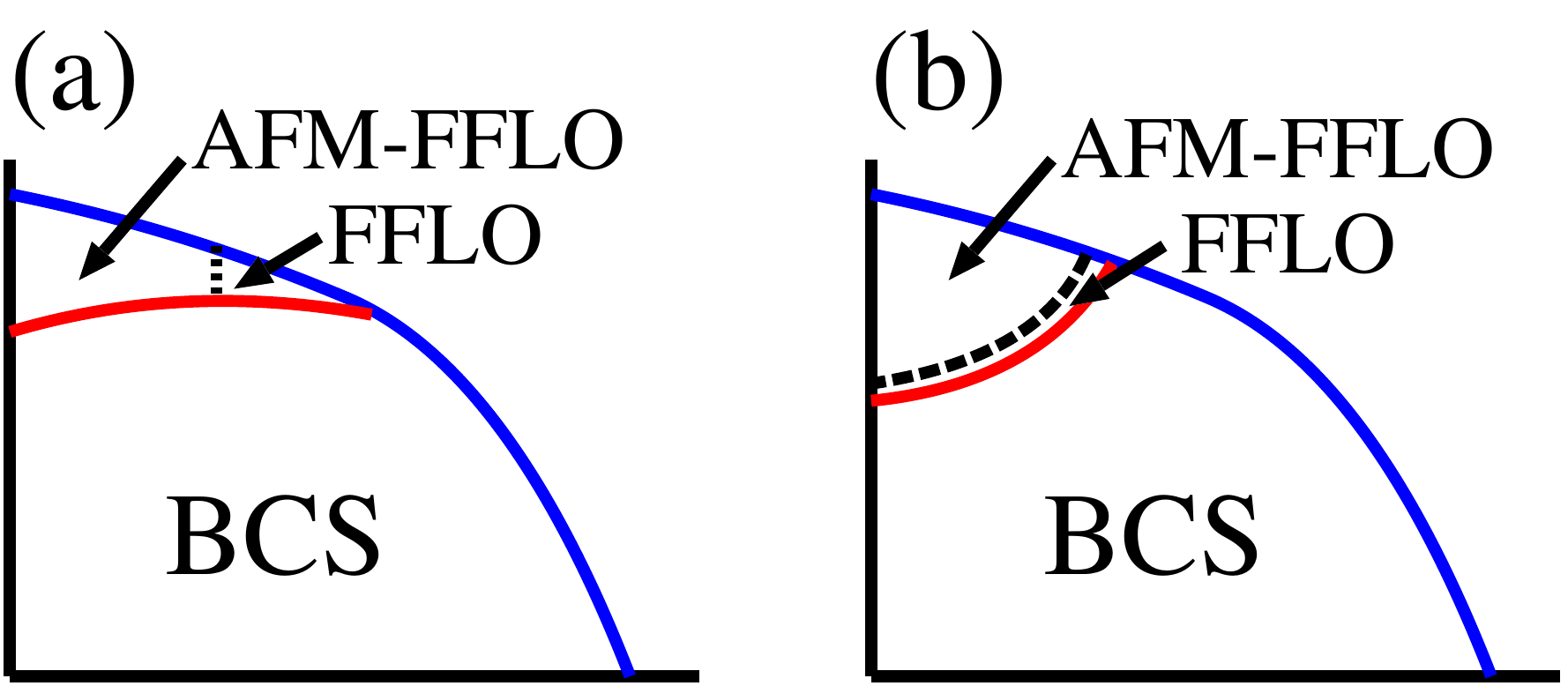}\hspace{2pc}%
\caption{(Color online)
(a) Schematic view of the $H$-$T$-phase diagram 
obtained by the mean field BdG equation in this paper. 
(b) Phase diagram in which the spin fluctuation is phenomenologically 
taken into account. 
The solid line in the superconducting state shows the BCS-FFLO transition, 
while the dashed line shows the Ne\'el temperature. 
}
\end{center}
\end{figure}

 We turn to the roles of the spin fluctuations neglected 
in the BdG equations. 
 First, the curvature of the BCS-FFLO transition line is changed by 
spin fluctuations near the AFM quantum critical point. 
 We have investigated the FFLO superconductivity in the two-dimensional 
Hubbard model on the basis of the FLEX approximation.  
The resulting BCS-FFLO transition line shows a concave 
curvature~\cite{yanaseFFLOQCP}, as observed for the HFSC phase 
of \Cof~\cite{miclea2006}. 
 This is mainly due to the quasiparticle lifetime renormalization
by the spin fluctuation that suppresses the FFLO state 
in the high-temperature region~\cite{yanaseFFLOQCP}. 
 This is in sharp contrast to the BdG equation that shows the convex 
BCS-FFLO transition line as in Fig.~1.

 Second, we discuss the mechanism that stabilizes the 
incommensurate AFM order in the FFLO state. 
 Spin susceptibility in the spatially uniform state is enhanced at 
$\vec{q}=\Q+\vdelta_{\rm IC}$ nearly independent of the direction of 
$\vdelta_{\rm IC}$ in the {\it ab}-plane on the tetragonal lattice. 
 Note that the directional fluctuations of  $\vdelta_{\rm IC}$ 
suppress the weakly incommensurate AFM order in the uniform phases, 
such as the normal and BCS states. 
 On the other hand, the appearance of the significant in-plane anisotropy 
due to the modulated FFLO order parameter suppresses 
the directional fluctuations and favors the AFM order. 
 This mechanism may play a role in stabilizing the AFM order in the 
HFSC phase of \Cof.

 Finally, the spin fluctuations play another role just above the 
BCS-FFLO transition line. 
 The continuous phase transition from the BCS state to the FFLO state is 
described by the nucleation of domain walls. 
 The density of domain walls approaches zero near the BCS-FFLO 
transition line. 
 Then, the spatial dimensionality of the AFM order is reduced from 
three to quasi-two-dimensions. 
 The reduced dimension enhances the fluctuations and suppresses the 
long-range order at finite temperatures when we neglect the broken 
SU(2)symmetry due to the spin-orbit coupling. 
 The AFM order is suppressed by this effect just above the BCS-FFLO transition 
line. 
 The density of domain walls rapidly increases with growing 
magnetic field from the BCS-FFLO transition line, and then, 
this effect of spin fluctuation is suppressed.

 Taking into account the roles of spin fluctuations, 
we obtain the schematic phase diagram in Fig.~6(b). 
 The AFM order is confined in the FFLO state, but it is suppressed 
around the BCS-FFLO transition line. 
 The pure FFLO state is stabilized just above the BCS-FFLO transition line. 
 The two phase transition lines, namely, the BCS-FFLO transition and the 
AFM order, are close to each other. 
 Therefore, it may be difficult to distinguish these transition lines 
in an experiment. 
 The \Co at ambient pressure seems to be the case.

\section{Summary and Discussion}

 We investigated the incommensurate AFM order in the $d$-wave spin singlet 
superconducting state on the basis of the BdG equations. 
 It has been shown that the AFM order coexists with the FFLO superconducting 
state even when the AFM order occurs neither in the normal state nor in the 
BCS state. 
 In other words, the AFM phase can be confined in the FFLO phase 
at high fields, consistent with the 
experimental results for \Cof~\cite{young2007,kenzelmann2008}. 
 Magnetic instability is enhanced in the inhomogeneous Larkin-Ovchinnikov 
state because the $\pi$-phase shift of the pairing field gives rise to the 
Andreev bound states and produces the large local DOS at zero energy. 
 The mixing with the $\pi$-triplet pairing also enhances the 
AFM order in the FFLO state.

 The structures of the AFM order, namely, the directions 
of the AFM magnetic moment $\vec{M}_{\rm AF}(\i)$ 
and incommensurability $\vdelta_{\rm IC}$, 
are consistent with the recent 
neutron scattering measurement for the HFSC phase of 
\Cof~\cite{kenzelmann2008}. 
 Both $\vec{M}_{\rm AF}(\i)$ and $\vdelta_{\rm IC}$ are 
perpendicular to the magnetic field. 
 It has been shown that the amplitude of the incommensurability 
$|\vdelta_{\rm IC}|$ is independent of the density of spatial nodes 
of the modulated superconducting order parameter~\cite{yanaseFFLOAF}. 
 This is also consistent with the experimental result~\cite{kenzelmann2008}. 
 These results indicate that the HFSC phase of \Co is the AFM-FFLO 
state in which the incommensurate AFM order coexists with 
the FFLO superconductivity. 
 To obtain unambiguous evidence for this scenario, we propose another 
neutron scattering experiment that can detect 
the spontaneous translation symmetry breaking along the magnetic field.

\begin{figure}[h]
\begin{center}
\hspace*{3mm}
\includegraphics[width=8cm]{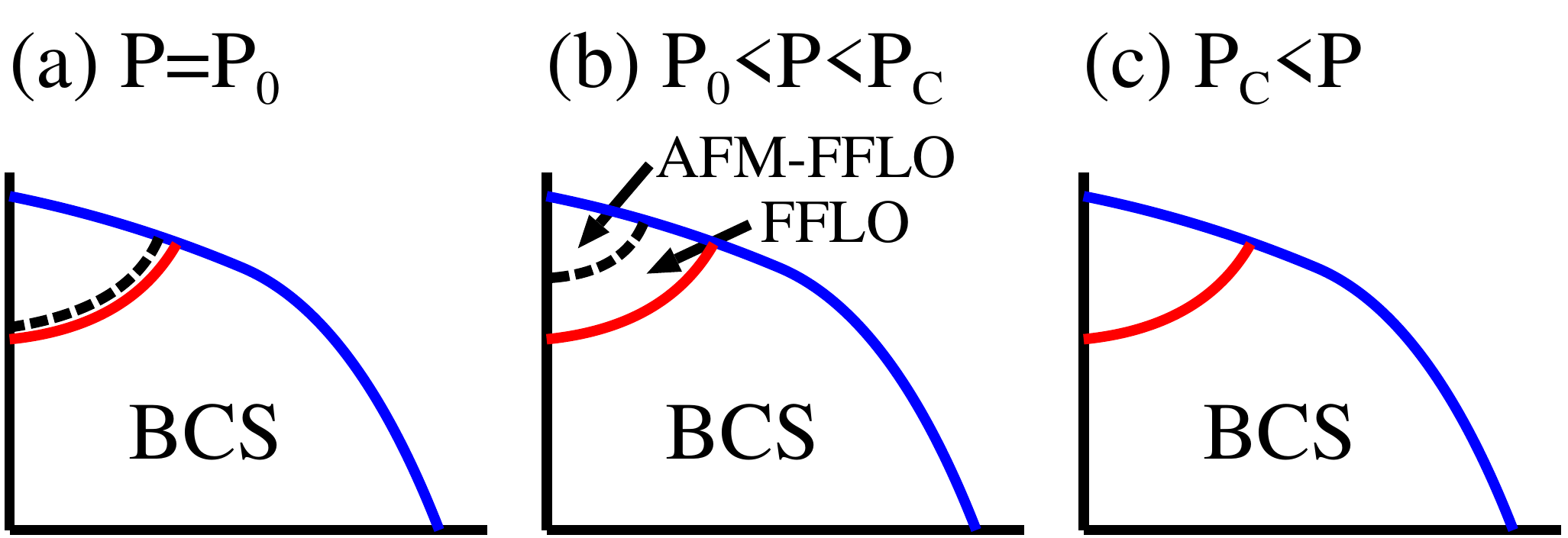}\hspace{2pc}%
\caption{(Color online)
Schematic figures of the $H$-$T$-phase diagram in \Co 
(a) at ambient pressure ($P = P_{0}$), (b) below the critical pressure 
($P_{0} < P < P_{\rm c}$), and (c) above the critical pressure ($P_{\rm c} < P$). 
}
\end{center}
\end{figure}

Observations under pressure may be a further way to explore 
the HFSC phase of \Cof. 
 Since the AFM order is expected to be suppressed by the pressure as in 
the other Ce-based heavy fermions~\cite{kitaoka2004}, the AFM-FFLO state 
is gradually suppressed by the pressure, as schematically shown in Fig.~7. 
 This is in sharp contrast to the pure FFLO state that is enhanced by 
the pressure as it goes away from the quantum critical 
point~\cite{yanaseFFLOQCP}. 
 Thus, the AFM order should be distinguished by the BCS-FFLO transition 
under pressure below $P < P_{\rm c}$, as shown in Fig.~7(b). 
 When the pressure exceeds the critical value $P_{\rm c}$, the AFM-FFLO phase 
vanishes, as shown in Fig.~7(c). 
 The experimental data for the pressure dependence are consistent with 
the enhanced FFLO phase~\cite{miclea2006}, 
but the magnetic order in the low-pressure region has not yet been studied.

 Finally, we discuss the experimental results for \Cof. 
 One of the key experiments is that on the pressure dependence of 
the $H$-$T$-phase diagram~\cite{miclea2006} mentioned above. 
 It has been shown that the HFSC phase of \Co is enhanced by 
the pressure. This experimental result is hardly understood by 
regarding the HFSC phase as an AFM ordered state 
in the uniform superconducting state or in the simple Abrikosov vortex state. 
 Therefore, the other quantum condensed state likely emerges in 
the HFSC phase of \Cof, such as the FFLO superconductivity 
investigated in this paper or the $\pi$-triplet pairing (pair density wave) 
proposed by other authors~\cite{aperis2008,aperis2009,agterberg2009}.


\section*{Acknowledgements}
 The authors are grateful to M. Ichioka, R. Ikeda, 
K. Ishida, M. Kenzelmann, K. Kumagai, K. Machida, Y. Matsuda, 
V. F. Mitrovi\'c, and K. Mizushima 
for fruitful discussions. 
This study has been supported by Grant-in-Aid for Scientific Research 
on Priority Areas "Superclean" (No.20029008), 
Grant-in-Aid for for Scientific Research on Innovative Areas 
"Heavy Electrons" (No.21102506), and Grant-in-Aid for 
Young Scientists (B) (No.20740187) from the MEXT, Japan, 
and by the Center for Theoretical Studies of ETH Zurich. 
 The numerical computation in this work was carried out 
at the Yukawa Institute Computer Facility.

\bibliographystyle{jpsj}
\bibliography{FFLOAF}

\end{document}